\documentstyle[preprint,psfig]{aastex}

\lefthead{Raga, Noriega-Crespo \& Vel\'azquez}
\righthead{X-ray luminosities of HH objects}

\def \yskip{\penalty-50\vskip3pt plus 3pt minus 2pt}
\def \pp{\par \yskip \noindent \hangindent .4in \hangafter 1}
\def \abc#1#2#3#4 {\pp#1, {\sl#2}, {\bf#3}, #4}
\def \blank {\lower 5pt\hbox to 0.75in{\hrulefill}}

\newfont{\rten}{cmr10}

\begin{document}

\title{The X-ray luminosities of Herbig-Haro objects}

\author{A. C. Raga\altaffilmark{1}, A. Noriega-Crespo\altaffilmark{2},
P. F. Vel\'azquez\altaffilmark{1}}

\altaffiltext{1}{Instituto de Ciencias Nucleares,
UNAM, Apdo. Postal 70-543, 04510 M\'exico, D. F., M\'exico,
email: raga@astroscu.unam.mx}
\altaffiltext{2}{SIRTF Science Center,
California Institute of Technology, Caltech 220-6, Pasadena,
CA 91125, USA, email: alberto@ipac.caltech.edu}

\begin{abstract}
The recent detection of X-ray emission from HH~2 and HH~154 with
the Chandra and XMM-Newton satellites (respectively) have opened up
an interesting, new observational possibility in the field of
Herbig-Haro objects. In order to be able to plan further X-ray
observations of other HH objects, it is now of interest to be able to
estimate their X-ray luminosities in order to choose which objects
to observe. This paper describes a simple, analytic model for predicting
the X-ray luminosity of a bow shock from the parameters of the flow
(i. e., the size of the bow shock, its velocity, and the pre-shock
density). The accuracy of the analytic model is analyzed through
a comparison with the predictions obtained from axisymmetric, gasdynamic
simulations of the leading working surface of an HH jet. We find that
our analytic model reproduces the observed X-ray luminosities of HH~2
and HH~154, and we propose that HH~80/81 is a good candidate for future
observations with Chandra.
\end{abstract}



\section{Introduction}

More than two decades ago, Ortolani \& D'Odorico (1980) detected the
UV emission of the Herbig-Haro object HH~1 with the International
Ultraviolet Explorer (IUE). This observation opened up the new
possibility of carrying out ultraviolet observations of HH objects, which
resulted in a large number of papers describing results obtained with
IUE (see, e.~g., Moro-Mart\'\i n et al. 1996), the Hopkins Ultraviolet
Telescope (Raymond et al. 1997) and the Hubble Space Telescope
(HST, Curiel et al. 1995).

Very recently, Pravdo et al. (2001) have reported Chandra observations
of HH~2 and Favata et al. (2002) have reported XMM-Newton observations
of HH~154 which are the first X-ray detections ever of HH objects.
Even though HH~2 and HH~154 are detected only in a marginal way,
these observations open
up the new possibility of analyzing the X-ray properties of HH objects.
This is an exciting development in observations of outflows from young
stars, because it gives us the possibility of detecting fast, non-radiative
shocks which could be associated with the outflows but have not been
previously detected. The results of future X-ray observations of
HH objects, however, will depend on whether or not other HH objects
are bright in the 0.1 - 10 keV window observed by Chandra.

In the present paper, we derive a simple, analytic model which gives
the X-ray luminosity of a bow shock as a function of the flow parameters
(\S 2). We then compare this analytic model with predictions obtained
from axisymmetric simulations of the leading working surface of a jet
(\S 3) in order to evaluate its accuracy. Finally, in \S 4 we compare
the luminosity predicted from our model with the HH~2 observations
of Pravdo et al. (2001), and suggest other objects which appear
to be good candidates for future Chandra observations.

\section{The analytic model}

A simple estimate of the X-ray luminosity of an HH bow shock can
be obtained as follows. We first assume that the X-ray luminosity is
dominated by the contribution of the free-free emission of hydrogen,
and that most of the free-free continuum photons come out in the
X-ray wavelength range. Then, the X-ray emission per unit volume
is given by the classical formula
\begin{equation}
\Lambda_{ff}={1.85\times 10^{-27}{\rm erg\,cm^{-3}\,s^{-1}}}
T^{1/2} n^2\,,
\label{lambda}
\end{equation}
where the temperature $T$ and the number density $n$ (of H ions
or of free electrons) are in cgs units (see, e.~g., Osterbrock 1989,
p. 53). In order to evaluate
the errors introduced by neglecting the line emission, and by assuming
that all of the free-free emission is emitted in the X-ray wavelength
range, we have compared the radiated energy per unit volume given by
equation (\ref{lambda}) with the one predicted using the Chianti dataset
and software (Dere et al. 2001,
which includes the line emission, computed under the assumption of
coronal ionization equilibrium) for the 0.3 to 10 keV photon energy
range. Through this comparison, one obtains differences of factors of
2.5, 3.0, 1.1 and 1.2 at temperatures of $10^5$, $10^6$, $10^{6.5}$
and $10^7$~K, respectively, between the two emission coefficients.

We now assume that the X-ray emitting region corresponds to the
head of the bow shock, in which the gas has a temperature and density
of the order of the on-axis post-shock values. From the
strong shock jump conditions, we then obtain
\begin{equation}
T\approx {1.5\times 10^5 \,\rm K}\left({v_{bs}\over {\rm 100\,km\,s^{-1}}}
\right)^2\,,
\label{t}
\end{equation}
\begin{equation}
n\approx 4\,n_0\,,
\label{n}
\end{equation}
where $v_{bs}$ is the velocity of the bow shock relative to the
downstream material, and $n_0$ is the pre-bow shock number density.

Finally, we need an estimate of the volume of the emitting region.
We will assume that the bow shock is created by a dense, approximately
spherical ``obstacle'' (which would in practice correspond to the
head of the HH jet) of radius $r_b$. For a non-radiative,
high Mach number, $\gamma=5/3$ flow, the on-axis stand-off distance
between the obstacle and the bow shock has a value $\Delta r\approx
0.2\,r_b$ (see Van Dyke \& Gordon 1959; Raga \& B\"ohm 1987).
For a radiative bow shock, however, the standoff distance
has a value $\Delta r\approx d_{cool}$, as has been frequently stated
in the literature, and tested in detail for bow shock flows
by Raga et al. (1997).

The emitting volume $V$ can then be calculated in an approximate
way as
\begin{equation}
V\approx {2\pi\over 3}{r_b}^3\left[\left(1+{\Delta r\over r_b}\right)^3
-1\right]\approx 2\pi {r_b}^2 \Delta r\,,
\label{v}
\end{equation}
where the second equality corresponds to a first order Taylor expansion
in $\Delta r/r_b$. The volume given by equation (\ref{v}) corresponds
to the volume limited by two hemispheres of radii $r_b$ and
$r_b+\Delta r$. The on-axis standoff distance then has to be
calculated as
\begin{equation}
\Delta r=\min[0.2\,r_b, d_{cool}]\,,
\end{equation}
where the cooling distance $d_{cool}$ can be computed with
the interpolation formula of Heathcote et al. (1998)
\begin{equation}
d_{cool}={\rm 2.24\times 10^{14}\,cm}\left({{\rm 100\,cm^{-3}}\over
n_0}\right)\left({v_{bs}\over {\rm 100\,km\,s^{-1}}}\right)^{4.5}\,,
\label{d}
\end{equation}
which fits the ``fully preionized'' plane parallel shock models
of Hartigan et al. (1987) in the $v_{bs}=150-400$~km~s$^{-1}$ shock
velocity range with $\sim 20$~\%\ accuracy.

Combining equations (\ref{lambda})-(\ref{d}), we then obtain
the following estimate of the X-ray luminosity $L_x$ of a bow shock~:
\begin{equation}
L_x\approx {\Lambda_{ff}\times V}=\min[L_r,L_{nr}]\,,
\label{lx}
\end{equation}
where
$$L_r={4.1\times 10^{-6}L_\odot}
\left({n_0\over {\rm 100\,cm^{-3}}}\right)\times$$
\begin{equation}
\left({r_b\over {\rm 10^{16}\,cm}}\right)^2\left({v_{bs}\over
{\rm 100\,km\,s^{-1}}}\right)^{5.5}\,,
\label{lr}
\end{equation}
$$L_{nr}={4.5\times 10^{-5}L_\odot}
\left({n_0\over {\rm 100\,cm^{-3}}}\right)^2\times$$
\begin{equation}
\left({r_b\over {\rm 10^{16}\,cm}}\right)^3\left({v_{bs}\over
{\rm 100\,km\,s^{-1}}}\right)\,.
\label{ln}
\end{equation}

\section{Numerical simulations}

In order to test the analytic model of \S 2,
we have carried out numerical simulations of HH jets, and
obtained predictions of the X-ray luminosities from the leading bow shock
which can be directly compared with the analytic model of \S 2.

In order to compute the gasdynamic jet simulations, we have used an
axisymmetric version of
the adaptive grid yguaz\'u-a code. This code has been described
in detail by Raga et al. (2000), and tested with ``starting jet'' laboratory
experiments by Raga et al. (2001). The version of the code that has
been used integrates rate equations for all of the ions of H and He,
and up to three times ionized C and O. The cooling processes associated
with these ions are included, as well as a parametrized cooling for
high temperatures. The details of the cooling functions and of the
ionization, recombination and charge exchange processes which are
included are given in the appendix of Raga et al. (2002).

The simulations have been carried out in a cylindrical, 4-level binary adaptive
grid with a $5\times 10^{18}$~cm (axial) and $6.25\times 10^{17}$ (radial)
spatial extent, and a maximum resolution of $4.88\times 10^{15}$~cm
along both axes. An initially top hat jet is injected on the left boundary
of the grid, and a reflection condition is applied outside the jet cross
section.

We have run three models, which share the following parameters. The
initially top-hat jet has a $r_j={5\times 10^{16}}$~cm radius,
number density $n_j=10^4$~cm$^{-3}$ and
temperature $T_j=1000$~K, and travels into a uniform environment
of density $n_{env}=400$~cm$^{-3}$ and temperature $T_{env}=1000$~K.
The jet-to-environment density ratio therefore has a $\beta^2=25$
value. Both the jet and the environment are initially neutral,
except for C, which is singly ionized.

The three jet models differ in their jet velocities~:
\begin{itemize}
\item M1~: this model has a $v_j=850$~km~s$^{-1}$ jet velocity. From
the usual ram pressure balance argument, one can calculate the on-axis
shock velocity of the leading bow shock as $v_{bs}=\beta v_j/(1+\beta)=
708$~km~s$^{-1}$ (where $\beta=\sqrt{n_j/n_{env}}=5$, see above).
For the parameters of this model, from equation (\ref{d}) we
obtain a $d_{cool}={3.7\times 10^{17}}{\rm \, cm}=7.5\,r_j$ cooling
distance. Therefore the stagnation region of the bow shock is
non-radiative.
\item M2~: this model has $v_j=360$~km~s$^{-1}$, resulting in
an on-axis shock velocity $v_{bs}=300$~km~s$^{-1}$ for the leading
bow shock and a $d_{cool}={1.5\times}$ ${10^{15}}$
${\rm \, cm}=0.03\,r_j$
cooling distance. Therefore, the head of the bow shock is radiative.
\item M3~: this model has $v_j=150$~km~s$^{-1}$, resulting in
an on-axis shock velocity $v_{bs}=125$~km~s$^{-1}$ for the leading
bow shock. For these parameters, equation (\ref{d}) gives a
$d_{cool}={1.3\times 10^{14}}{\rm \, cm}=0.0025\,r_j$ cooling distance.
Because the shock velocity of the model is outside the range of
validity of equation (\ref{d}), this cooling distance differs by a factor
of $\sim 3$ from the one given by Hartigan et al. (1987).
\end{itemize}
We should note that the cooling distances of the last two models are not
appropriately resolved in our numerical simulations.

As an example of the results obtained from the
computed jet models, in Figure 1 we show the density stratification of
the jet head obtained from model M1 for a $t=2000$~yr integration time.
From this Figure, we see that the dense, post-Mach disk jet material
is ejected sideways, forming a cocoon which constitutes
the ``obstacle'' which pushes the leading bow shock into the surrounding
environment. Therefore, in order to compare the numerical simulations
with the analytic model of \S 2, we associate the ``radius of the
obstacle'' $r_b$ (see equation \ref{v}) with the cylindrical radius
of the dense cocoon.
We find that for the three computed models, the radius
of the cocoon has oscillations of $\sim 20$\%\ around a $r_b=
10^{17}$~cm value. We therefore adopt this value in order to
compare the X-ray luminosities predicted from the numerical simulations
with the analytic model of \S 2.

Figure 2 shows the time-dependent X-ray luminosity of the heads
of the jets computed from models M1, M2 and M3. The luminosities
have been computed as follows. The frequency-dependent emission coefficient
has been calculated using the Chianti data set and software
(Dere et al. 2001), under the assumption of coronal ionization
equilibrium and with solar abundances. This emission coefficient has
then been integrated over energies ranging from 0.3 to 10 keV (in
order to simulate Chandra observations), and over all of the volume
of the computational domains of the jet models.

In Figure 2, we also show the values of $L_x$ obtained from
our analytic model (see equations \ref{lx}-\ref{ln}). For models
M1 and M2, we obtain good agreement (within a factor of $\sim 2$) between
the predictions from the numerical simulations and the analytic model.

For model M3, the $L_x$ values obtained from
the numerical simulation range from a factor of $\sim 4$
to a factor of $\sim 10$ times the analytic prediction (see figure 2).
This larger difference between the numerical and analytic predictions
might be due to the fact that the cooling distance of model M3 is
highly unresolved (see above).

We then conclude that the X-ray luminosities obtained from the
numerical simulations and from the analytic model are
in reasonably good agreement.

\section{The X-ray luminosities of three bright HH objects}

\subsection{HH~2}

This object has been marginally detected with Chandra by Pravdo
et al. (2001), who estimated a (redenning corrected) X-ray luminosity
of $L_x\approx {1.3\times 10^{-4}}L_\odot$. The detected
emission is associated with the HH~2H condensation.

HST images of HH~2H (Schwartz et al. 1993) show that this condensation
has a high intensity, elongated structure extending more or less
perpendicular to the outflow axis. The lateral extension of this
structure is of $\approx 2''.5$, which corresponds to a physical
size of $\approx {1.7\times 10^{16}}$~cm (at a distance of 460~pc).
We therefore have $r_b\approx {8.5\times 10^{15}}$~cm.

For the shock velocity and pre-shock density, we adopt the
values estimated for
HH~2H by Hartigan et al. (1987). Following these authors, we
set $v_{bs}=150$~km~s$^{-1}$ and $n_0=500$~cm$^{-3}$.

From equation (\ref{lr}) we then obtain $L_x={1.4\times 10^{-4}}L_\odot$.
Therefore, the prediction obtained from our analytic model is
in surprisingly good agreement with the HH~2H
luminosity determined by Pravdo et al. (2001).

\subsection{HH 154}

HH~154 is a chain of aligned knots leading away from the L~1551 IRS~5
source. Favata et al. (2002) deduce that the X-ray emission that
they detect comes from the high excitation knot D, for which they
deduce (from previously published optical line ratios and line profiles)
a $v_{bs}\approx 200$~km~s$^{-1}$ bow shock velocity.
Favata et al. (2002) and Fridlund \& Lizeau (1994, 1998) deduce
an $\approx 500$~cm$^{-3}$ density for the region upstream of knot D,
and estimate that the density downstream of the knot has to
be lower than this value by a factor of $\sim 15$ (taking the
mean of the 10-20 range quoted by Fridlund \& Lizeau 1998 for
this factor). Therefore, we set $n_0\approx 100$~cm$^{-3}$.
Finally, from the HST images of Fridlund \& Lizeau (1998), we see that
knot D has a diameter of $\approx 3''.3$, corresponding to
$r_b={7.5\times 10^{15}}$~cm (at a distance of 150~pc).

With these parameters, from equation (\ref{lr}) we obtain
$L_x={1.0\times 10^{-4}}L_\odot$. This luminosity is in uncannilly
good agreement with the ${8\times 10^{-5}}L_\odot$ X-ray luminosity
which Favata et al. (2002) deduced from their XMM-Newton observations.

\subsection{HH 80/81}

The HH objects with the highest excitation spectra are HH~80 and 81.
These objects (discovered by Reipurth \& Graham 1988) are associated
with a thermal radio jet which shows proper motions of up to
$\sim 1400$~km~s$^{-1}$ (Mart\'\i\ et al. 1998).

HST images (Heathcote et al. 1998)
show that HH~81 has an angular size of $\approx 3''.5$,
corresponding to a physical size of $\approx {8.9\times 10^{16}}$~cm
(at a distance of 1700~pc). We therefore adopt
$r_b={4.5\times 10^{16}}$~cm.
We also adopt the $v_{bs}\approx 700$~km~s$^{-1}$ and
$n_0\approx 400$~cm$^{-3}$ values deduced from the line widths
and H$\alpha$ luminosity of HH~81 by Heathcote et al. (1988).
Using these values, from equation (\ref{ln}) we obtain
$L_x=0.46\,L_\odot$.

In order to estimate whether or not this object can be detected
with Chandra, we have to calculate the energy flux $F_{HH~81}$
that would arrive at Earth. We use the distance of 1700~pc
and the $A_V=2.33$ (corresponding to a $N_H\approx
{3.0\times 10^{21}}$~cm$^{-2}$ neutral H column density) determined by
Heathcote et al. (1998). If we consider the extinction
at 1~keV, we obtain $F_{HH81}\approx
{2.5\times 10^{-12}}$~erg~s$^{-1}$~cm$^{-2}$, and if we use the
extinction at 0.5~keV, we obtain  $F_{HH81}\approx
{7.5\times 10^{-15}}$~erg~s$^{-1}$~cm$^{-2}$

Interestingly, the lower estimate that we have obtained
for the flux that would be observed from HH~81 is an order of magnitude
higher than the flux observed by Pravdo et al. (2001) for HH~2.
Therefore, we conclude that HH~81 is a very good candidate for
future Chandra observations.

\section{Conclusions}

We have derived a simple, analytic model for predicting the X-ray
luminosity of HH bow shocks. We have tested this model against
HH jet numerical simulations, showing that it is applicable for
bow shocks with shock velocities in the $120\to 600$~km~s$^{-1}$ range.

We have then applied the analytic model to obtain predictions of
the X-ray luminosities of HH~2H, HH~154D and HH~81, and obtain the following
results~:
\begin{itemize}
\item the luminosity predicted for HH~2H is in good agreement with
the Chandra observation of this object by Pravdo et al. (2001),
\item the luminosity predicted for HH~154D is in good agreement
with the XMM-Newton observation of this object by Favata et al. (2002),
\item the X-ray luminosity predicted for HH~81 is a factor of
$\sim 1000$ larger than the one of HH~2.
\end{itemize}
Because of the very large difference in X-ray luminosities predicted
for HH~81 and for HH~2, even though HH~81 is a factor of $\sim 3$ more
distant and more highly extinguished than HH~2, we expect it to
be brighter than HH~2 by a factor of 10 to 100.

\acknowledgments
The work of AR and PV was supported by the CONACyT grants 34566-E and
36572-E. ANC's research was carried out at the Jet Propulsion Laboratory,
California Institute of Technology, under a contract with NASA;
and partially supported by NASA-APD Grant NRA0001-ADP-096.

\begin{figure}
\psfig{figure=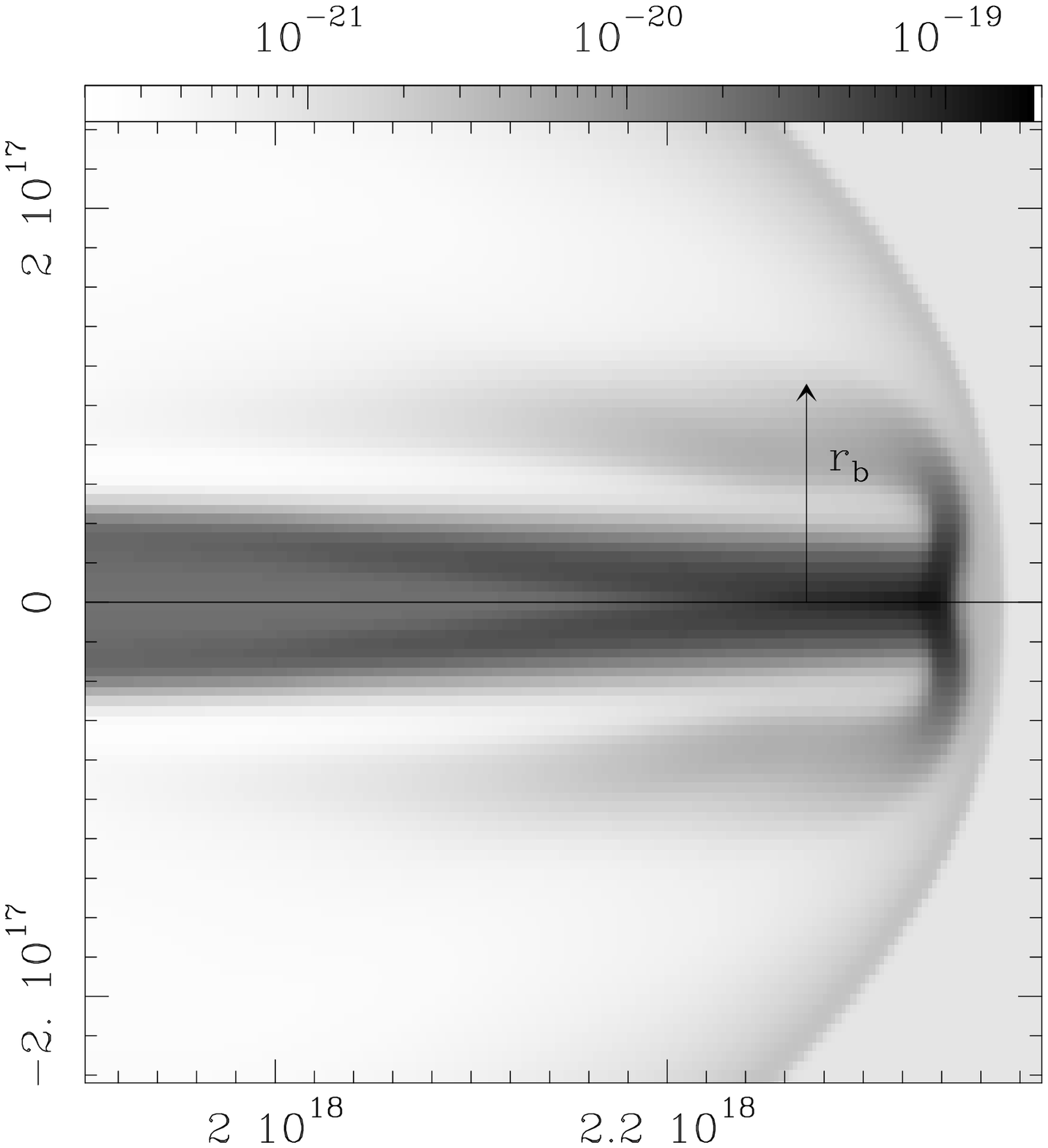,width=7.8cm}
\figcaption{Density stratification of the region around the head
of the jet obtained from model M1 for a $t=2000$~yr time-integration.
The outer radius $r_b$ of the dense cocoon formed by the post-Mach disk
jet material is indicated on the figure. The density is shown with
a logarithmic greyscale, given by the bar on the top of the plot
in g~cm$^{-3}$. The two axes are labeled in cm, with the zero point
corresponding to the centre of the initial jet cross section.
\label{fig1}}
\end{figure}

\begin{figure}
\psfig{figure=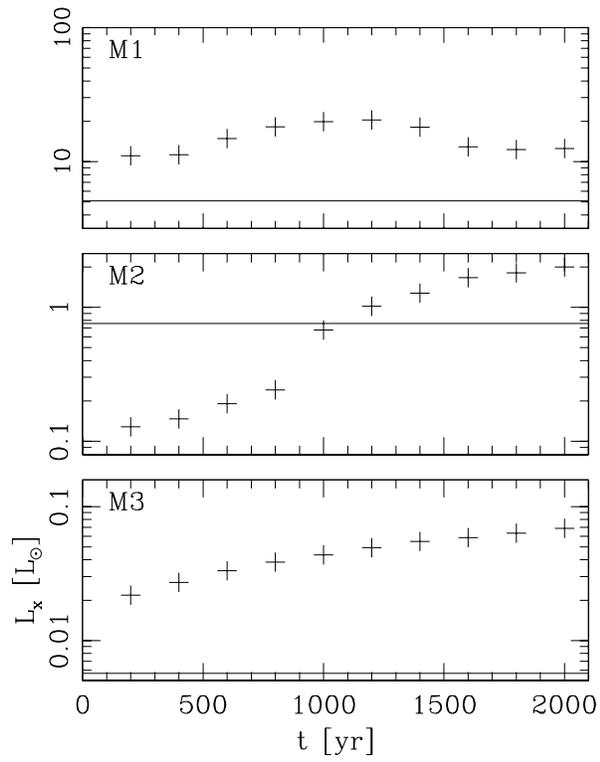,width=7.8cm}
\figcaption{X-ray luminosity in the 0.3-10~keV energy range computed
from models M1, M2 and M3 as a function of integration time. The horizontal
lines represent the (time-independent) X-ray luminosity predicted for
the appropriate parameters from the analytic model described in \S 2.
\label{fig2}}
\end{figure}

\end{document}